**A Survey Study to Understand Industry Vision for Virtual and Augmented Reality Applications in Design and Construction**

Mojtaba Noghabaei, S.M.ASCE[1]; Arsalan Heydarian, A.M.ASCE[2]; Vahid Balali, A.M.ASCE[3]; and Kevin Han, A.M.ASCE[4]

[1]PhD Student, Department of Civil, Construction, and Environmental Engineering, North Carolina State University, Raleigh, NC 27695; PH (919) 798-6820; email: snoghab@ncsu.edu

[2]Assistant Professor, Department of Civil and Environmental Engineering, Link Lab, University of Virginia, Charlottesville, VA 22903; PH (434) 924-1014; email: heydarian@virginia.edu

[3]Assistant Professor, Department of Civil Engineering and Construction Management, California State University, Long Beach, CA 90840; PH (562) 985-1643; email: vahid.balali@csulb.edu

[4]Assistant Professor, Department of Civil, Construction, and Environmental Engineering, North Carolina State University, Raleigh, NC 27695; PH (919) 515-8719; email: kevin_han@ncsu.edu

# Abstract

With advances in Building Information Modeling (BIM), Virtual Reality (VR) and Augmented Reality (AR) technologies have many potential applications in the Architecture, Engineering, and Construction (AEC) industry. However, the AEC industry, relative to other industries, has been slow in adopting AR/VR technologies, partly due to lack of feasibility studies examining the actual cost of implementation versus an increase in profit. The main objectives of this paper are to understand the industry trends in adopting AR/VR technologies and identifying gaps between AEC research and industry practices. The identified gaps can lead to opportunities for developing new tools and finding new use cases. To achieve these goals, two rounds of a survey at two different time periods (a year apart) were conducted. Responses from 158 industry experts and researchers were analyzed to assess the current state, growth, and saving opportunities for AR/VR technologies for the AEC industry. The authors used t-test for hypothesis testing. The findings show a significant increase in AR/VR utilization in the AEC industry over the past year from 2017 to 2018. The industry experts also anticipate strong growth in the use of AR/VR technologies over the next 5 to 10 years.

*Keywords*: Virtual Reality; Augmented Reality; Building Information Modeling; Industry Trend

# Introduction

One of the largest industries in the United States is the AEC industry with expenditure reaching over $1.162 trillion in 2017 (Statista 2017). Over 98% of construction projects incur cost overruns and delays (Changali et al. 2015). Many projects experience rework, costing 5% to 20% of the total contract value (Forcada et al. 2017). The main causes of rework include lack of communication among different construction parties, lack of adequate visualization capability to recognize design conflicts, and lack of support for advanced communication technologies (Fayek et al. 2003; Gamil and Abdul Rahman 2018). Addressing these deficiencies can decrease the number of unforeseen issues and, therefore, rework in construction projects (Gamil and Abdul Rahman 2018).

Over the past decade, BIM has found a wide range of applications in the AEC industry (Noghabaei et al. 2019, 2020; Noghabaei and Han 2020; O'Neill et al. 2020; Sherafat et al. 2019b; a, 2020; Volk et al. 2014; Yalcinkaya and Singh 2015). In this paper, BIM is defined as the process of generating and involving digital representation of a building or construction and their characteristics. BIM is not just the production of 3D models (Chong et al. 2016) therefore it can be used for different functions such as, improving communication, decision making enhancement, and visualization. Furthermore, BIM can accelerate information integration form design to construction (Liao and Ai Lin Teo 2018). BIM technology has improved and revolutionized the way designers, engineers, and managers think about the buildings and enables them to predict and solve problems that might occur during the life-cycle of a building. BIM technology has enabled designers and engineers to detect clashes and simulate different construction scenarios for more efficient decision making. BIM technology revolutionized the AEC industry in many different aspects, such as technical aspects, knowledge management, standardization, and diversity management (Ghaffarianhoseini et al. 2017). However, BIM still has some inherent shortcomings. For instance, BIM does not provide robust visualization for cluttered construction sites and the existing software packages provide limited user experiences (i.e., lack of interactive visualization using a keyboard and mouse) (Du et al. 2018). In addition, the other challenge in many construction projects lies within interoperability of BIM tools and IFC (McCuen et al. 2012). Moreover, investigations have shown that BIM has some limitations in real-time on-site communication (Wang et al. 2013). Additionally, the stakeholders who are not familiar with BIM solutions are not able to utilize its capabilities, such as improved communication through visualization. Finally, BIM does not provide sense of realism while immersive virtual mock-ups with a one-to-one scale can create improved interaction with BIM which results in improved communication at the end (Heydarian et al. 2015).

To address some of the inherent deficiencies of BIM, researchers proposed the use of new technologies such as Augmented Reality (AR) and Virtual Reality (VR). In this paper, AR is referred to a physical environment, whose elements are augmented with and supported by virtual input and VR is referred to a simulated virtual environment, representing a physical environment. Accordingly, Immersive Virtual Environments (IVEs) are environments where user interaction is supported within a virtual environment. AR/VR technologies can potentially address these deficiencies and enhance BIM in several aspects, such as real-time on-site communication (Wang et al. 2013). AR/VR can also improve communication among stakeholders and provide better visualization for engineers, designers, and other stakeholders, enabling one-to-one fully immersive experience (Biocca and Levy 2013). Furthermore, IVEs have the necessary potentials to achieve knowledge synthesis to improve design process (Dossick et al. 2015). Despite all these benefits, the AEC industry is far behind other industries in adopting AR/VR technologies.

Many industries implemented AR/VR in a successful way. For example, AR/VR has applications in manufacturing (Berg and Vance 2017; Choi et al. 2015), retail (Bonetti et al. 2018; Dacko 2017), mining (Pedram et al. 2017; Zhang 2017), education (Greenwald et al. 2017; Merchant et al. 2014; Zhang et al. 2018) and healthcare, especially for simulating surgeries (Atwal et al. 2014; Khor et al. 2016; de Ribaupierre et al. 2014). Recent studies indicate the benefits of AR/VR in the AEC industry by demonstrating potential applications, such as safety training (Li et al. 2018), visualization (Fogarty et al. 2018; Paes et al. 2017), communication (Du et al. 2018), and energy management (Niu et al. 2016).

Although research suggests AR/VR technologies can be very effective, the AEC industry has been very slow in adopting these technologies, which could be partly due to lack of feasibility, examining the actual cost of implementation versus an increase in profit.

The main objective of this study is to understand the industry trends (growth pace) and limitations in adopting AR/VR technologies, as well as identifying new opportunities for leveraging AR/VR. Through a set of comprehensive survey, the authors investigated the utilization growth of AR/VR technologies within the AEC industry. Furthermore, this paper aims to understand potential cost and time savings and find opportunities for AR/VR developments in order to improve communication and visualization among different stakeholders. To achieve these objectives, a series of two surveys over a one year period is conducted where over 150 AEC industry experts have provided their feedbacks and visions on growth and utilization of AR/VR technologies within the AEC industry.

## Literature Review

In this section, the authors investigated the potentials and applications of AR/VR technologies in AEC and other domains such as, education, healthcare, mining industry, and retail industry. This comparison between AEC and other domains shows potential use cases of AR/VR in AEC industry.

### AR/VR in Other Domains

Over the past decade, many researchers in different fields have investigated how AR/VR tools can enhance the communication of information among users. For instance, in the retail industry, Javornik (2016) and McCormick et al. (2014) demonstrated that AR/VR applications are rapidly evolving and increasingly used over the past years. Dacko (2017) quantitatively analyzed more than 250 Mobile Augmented Reality (MAR) applications for shopping. The results demonstrated that MAR is beneficial (i.e., efficiency or better shopping value) to the retail industry and presented actions to leverage MAR for smart retail.

In addition to the aforementioned industries, the mining industry is one of the pioneer industries in adopting AR/VR technologies. Grabowski and Jankowski (2015) demonstrate that a VR solution can enhance occupational health and safety of coal mining workers by presenting a pilot study. In this study, the workers were trained by professionals who had adequate experience with safety training. They tested different motion capture systems, Head-Mounted Displays (HMD), joysticks as input methods, and training scenarios and compared the results. The results showed that VR technology can be a very effective platform, substitute on-site training, and prevent trainees from exposure to dangers and risks that are common in a mining environment. Zhang (2017) developed a VR-based training system for the mining industry and demonstrated that having more immersion using devices like magic leap can improve the training systems. Pedram et al. (2017) evaluated the VR-based safety training systems and concluded these systems have a positive learning experience.

AR/VR technologies have been receiving much attention in the healthcare industry due to their immersion capabilities. Mosadeghi et al. (2016) conducted a case study with over 500 hospital patients. The patients viewed VR simulations such as ocean exploration and tour of Iceland to reduce the stress level. Then, they conducted a survey on anxiety and pain level. The results demonstrated that most of the inpatient users expressed that VR experience was pleasant and it was capable of reducing pain and anxiety. Tashjian et al. (2017) designed a similar experiment with 50 patients. Patients viewed a 15-minute VR simulation called Pain RelieVR. This simulation designed in a way that can reduce the stress through a game-like experience. They monitored the heart rate and blood pressure of the patients during the experiment. The results of this experiment indicate that VR can significantly reduce pain versus traditional control distraction condition. Dascal et al. (2017) reviewed the applications of VR in healthcare industry between 2005 and 2015 and concluded that VR has shown more success in three areas: eating disorders, pain management, and cognitive and motor rehabilitation. In addition, Pelargos et al. (2017) investigated the potentials of using AR/VR in neurosurgery. They concluded that healthcare industry needs more AR/VR tools for educational purposes.

There are also many researchers in education who have investigated AR/VR technology. Akçayır and Akçayır (2017) presented a comprehensive review of usage, challenges, and advantages of AR technology in the education industry. They determined that AR can enhance learning achievements and motivate students. Potkonjak et al. (2016) show the growth in online education and distant-learning that uses IVE. Wei et al. (2015) developed an AR-based teaching system. They showed that teaching using their AR-based application increases student motivation and improves the innovation and creativity of the design outputs in a design course. Nikolic et al. (2009) developed a VR-based tool that is proved to be a reliable and effective solution to the challenges faced by students in visualizing 3D structures. It allows students to visualize and review various designs through a VR environment. The efficiency and usefulness of the tool were assessed by surveys, group interviews, and in-class exercises. The results showed that subjects had a far better understanding of concepts when using a VR interface.

**AR/VR in AEC**

Usage of AR/VR technologies in other fields such as healthcare, education, and retail has shown to be useful for improving human behavior, student learning enhancement, increasing revenues in retailing. The other fields are growing in this area and also recently, AEC has grown too, but more in some specific areas and not across the entire industry mainly because of lack of budget in the industry and as a result, AEC industry has not adopted these tools, but it is possible to improve budget and enhance scheduling if AR/VR are effectively used.

Utilization of IVEs in an engaging experience for end-users in project design process, and combining IVEs sense of presence and BIM models can enhance the opportunity to evaluate different alternative design options in a time and cost efficient approach.

The AEC industry has many potential use cases for utilizing AR/VR technologies such as, safety training, improving BIM visualization and communication, BIM-based immersive tools, energy savings, and understanding end-users (occupants) preferences. Li (2018) performed a case-study on personalized safety training in an IVE in order to achieve more efficient safety training with better results. Sacks et al. (2013) conducted a research study to evaluate the long-term effect of VR safety training in comparison to the traditional approaches. They performed an experiment with two groups of 30 respondents. They gave a VR-based training to the first group while the second group has gone through the traditional safety training program. The results of the study indicated that the VR-based safety training program is significantly more effective than the traditional approach in both short term and long term. Le et al. (2015) developed an online VR framework that enables workers to perform dialogic learning, role-playing, and social interaction to provide better safety and health education for the workers. They concluded that the platform effectively improves health and safety education. Jeelani et al. (2017) developed a training strategy that simulated construction accidents in the VR environment to demonstrate accident causation and the importance of thorough hazard recognition and proper risk perception. After training, the workers were able to identify more hazards, perceive them with a higher level of risk and were able to use effective management strategies to control the hazards concluding that VR environments provide a high degree of realism which improves training outcomes.

Balali et al. (2018) developed a framework for cost estimation in construction using VR technology. They used a real-time VR model that can give the stakeholders and the users the ability to change the material of the walls, floors and other parts and the model provides them the price impact in real-time. Linking cost estimation to VR can be beneficial to the AEC industry, especially to estimators. Du et al. (2018) introduced a cloud-based VR system called CoVR to improve communication among stakeholders in a construction project. CoVR is able to import BIM data and visualize it in a multiuser interactive virtual environment. This platform enables remote stakeholders to have social and face-to-face interaction with others. The researchers conducted a survey on CoVR and the results demonstrated that CoVR can enhance communication. Williams et al. (2015) developed an MAR application that can augment BIM models on top of the real world building. This application has the potentials to help technicians to optimize and visualize their model and data promptly in an AR application.

Some researchers used IVE to develop interactive training environment for workers, technicians, and engineers. Goulding et al. (2012) introduced a VR-based interactive environment that enables a user to interact with triggered problems on a construction site and make decisions. They can see how their decisions affect project cost and schedule. The respondents of this study were interested in the tools and believed that VR provided better interaction and improved the decision making. Fang and Cho (2016) developed a virtual prototyping platform to improve crane safety. In this platform, a lift crew, consisting of a planner, rigger, signalman, and operators, virtually perform lifting operations. The results indicate that this tool can improve the operator's confidence and safety. Kayhani et al. (2018) developed a VR platform that can simulate the heavy mobile crane lift in a modular construction. This platform enables the lift crew and engineers simulate the lift in an IVE and evaluate different options in real-time. This platform can simplify the heavy lift planning, improve the lift crew's performance on the construction site, and reduce human error.

Furthermore, some studies used IVE for improving the degree of presence in lighting condition assessment and energy management (Birt et al. 2017; Kuliga et al. 2015). Niu et al. (2016) developed a design approach combining VR and design with an intent concept that can help in closing energy performance gap caused by occupants' behavior. The results indicate that the developed framework can help designers detect design patterns that can predict actual occupant behaviors. Heydarian et al. (2015b) conducted an experiment to compare the respondents' sense of presence in a VR environment versus a real environment. A realistic model of a room with different lighting options was created. The respondents selected similar options in VR versus real room. The results showed that VR is effective in obtaining user feedback. The feedback can improve the end-user satisfaction rate and performance in design (Heydarian et al. 2017). In another IVE study, Heydarian et al. (2016) evaluated how psychological factors such as defaults and personality traits may influence occupant's lighting and shading interactions; through collecting data from over 150 participants, they concluded that without any additional cost, defaults can be used to significantly reduce the lighting electricity consumption in commercial buildings.

## Method

Since the implementation of AR/VR technologies is still relatively new in AEC industry, there is not much empirical data on these topics. In order to gather data, the authors came up with a number of research methods. First, the authors designed a detailed online questionnaire. The detailed questionnaire was reviewed by three BIM specialists as well as three researchers within the field of construction engineering and management to ensure questions are clear and not misleading. The authors designed the questionnaire in a way to be able to analyze the growth of these technologies by collecting responses at two different time periods. Finally, by further analysis on the survey results, the authors were determine the industry trends and visions over AR/VR technologies.

The questionnaire is formulated to gather the information about the AEC industry's adoption of AR/VR technologies over the past year. Moreover, the questionnaire investigated the opportunities for AR/VR technologies to improve stakeholders' communication and identify experts' predicted return on investment. The questionnaire is designed to analyze the growth of these technologies by collecting responses at two different time periods – once in spring 2017 and again in spring 2018. The online surveys were hosted on https://new.qualtrics.com/. Qualtrics enabled the authors to keep a record of the computer address from which the survey was completed using internet protocol (IP) and assign an identification number (ID) to the user's IP. Qualtrics excluded duplicated data by checking respondents' profiles, IPs, IDs, and entries from database for analyzing survey results. The excluded responses were mainly from the respondents who didn't complete the survey so that the authors could not accredit their credibility for the goals of this research.

As a first step, a set of 27 survey questions were designed to target a range of AEC professionals, such as engineers, designers, researchers, managers, and owners. The survey questions were divided into five sections: 1) general information, 2) company related information, 3) BIM knowledge, 4) general AR/VR related information, and 5) visions for future of AR/VR. The first three sections capture the background and experience of the respondents. Then, AR/VR is evaluated in the next two sections. Table 1. Description of target areas and objectives with respect to different parts of the survey describes the main sections, gathered data, and the objective of each section in more detail.

**Table 1.** Description of target areas and objectives with respect to different parts of the survey

| General Section | Section Name | Gathered Data | Objectives |
|---|---|---|---|
| background and experience | general information | age, gender, occupation, and professional experience | determine how respondents in different parties envision the future of AR/VR |

|  | company related information | companies size, turnovers, and employees number | assess how companies with different sizes envision the future of AR/VR |
| --- | --- | --- | --- |
|  | BIM knowledge and experience | BIM experience and used BIM tools | evaluate how respondents with different BIM knowledge envision the future of AR/VR |
| AR/VR evaluation | AR/VR knowledge and experience | AR/VR experience and used AR/VR tools | shows the industry trends |
|  | visions for the future of AR/VR | opportunities of AR/VR in the AEC industry | represents the industry future on AR/VR |

The first and second rounds of the survey had 94 and 64 respondents, respectively. The surveys were distributed directly among professionals within the AEC industry and also through the Construction Management Association of America (CMAA) organization. CMAA was chosen since it is has a great combination of 16000 members in AEC industry from both public and private sectors across the USA. CMAA expert members are from different parties such as owners, architects and designers, general contractors, and construction managers. Since CMAA has this wide range of members from different parties, the authors decided to use CMAA for distributing the surveys. The authors conducted the first round of survey in March through May 2017 and the second round in January through March 2018. The surveys were distributed in two rounds to measure the impact and growth of AR/VR in the AEC industry and enable the realistic evaluation of the current state of AR/VR.

The first section of the survey attempts to identify the general information of the respondents, such as age, gender, occupation, and professional experience. In the next section, the respondents answer several questions about their companies, such as geographical location, size, and type of projects (e.g., residential commercial, institutional, etc.). The third section examines the respondents' competency in BIM technology and applications (i.e. quality control, progress monitoring).

In the next two sections, the survey results assessing AR/VR utilization in the AEC industry as well as the future opportunities for AR/VR applications are presented. First, the respondents are asked what types of AR/VR devices they have used and how many AR/VR experts they have in their companies.

Through these questions, the authors were able to evaluate the respondents' familiarity with AR/VR tools and their companies' effort in integrating these technologies with on-going and future projects. In the last section, the respondents were asked to answer a few questions about their vision on the future integration of AR/VR technologies within the AEC industry. The questions in this section were designed in a way that demonstrates AR/VR potentials for future developments. For example, the respondents were asked to identify the sectors (i.e., education and healthcare facilities) and the project size that can best leverage AR/VR technology. The last section evaluated the visions for cost and time saving through integrating AR/VR technologies in construction projects. The last two questions evaluate how the respondents predicted the increase in end-users satisfaction when AR/VR technology is used and their limitations in AEC-related applications. By understanding the potential and maturity of AR/VR technologies, industry leaders can better understand the potential use-case of these tools. The identified industry trends can help industry leaders make better investment decisions on these technologies.

## Survey Findings and Results

In this section, the survey responses are analyzed to (1) understand the current state and growth of AR/VR in the AEC industry over the past year, (2) identify opportunities of AR/VR development in improving communication and visualization, and (3) understand the benefits, that are foreseen by AEC practitioners, of adopting AR/VR technologies.

In order to account for participant privacy, the surveys did not ask for any personal information such as, name, company name, and etc. from the participants. To detect whether participants took part in both rounds, the authors added a question to the second survey asking the participants whether they had participated in the same survey study previously. The results for this question demonstrated that none of the participants in the second round of survey participated in the first round. The survey results are analyzed as follow to understand these trends.

## General Respondent Information

Overall in both surveys, 71% (67% and 77% respectively in each survey) of the respondents were male and 29% were female (33% and 23% respectively in each survey). Respondent's age ranged from 25 to 60 with the average of 32 overall in both surveys. Approximately, 70% of the respondents (78 out of 114 respondents who were willing to share their age) were 30 years old or younger. Respondents were also asked about their roles in the AEC industry. The survey had four groups and the result is summarized in Table 22.

**Table 2.** Survey result: participants role in the AEC industry

|  | Owner | Manager | Researcher | Engineer and Designer |
|---|---|---|---|---|
| Survey 1 | 1 % | 29% | 21% | 49% |
| Survey 2 | 0 % | 17% | 32% | 51% |

The professional experience is another important indicator of the expertise of the respondents. Most of the respondents with expertise in BIM and AR/VR technologies were relatively young. Overall in both surveys, approximately 75% of the respondents indicated that they had 10 years or less of professional experience in the AEC industry. Fig. 1 shows the number of years the respondents have spent at their current companies and Fig. 2 presents how many years they have worked in the AEC industry.

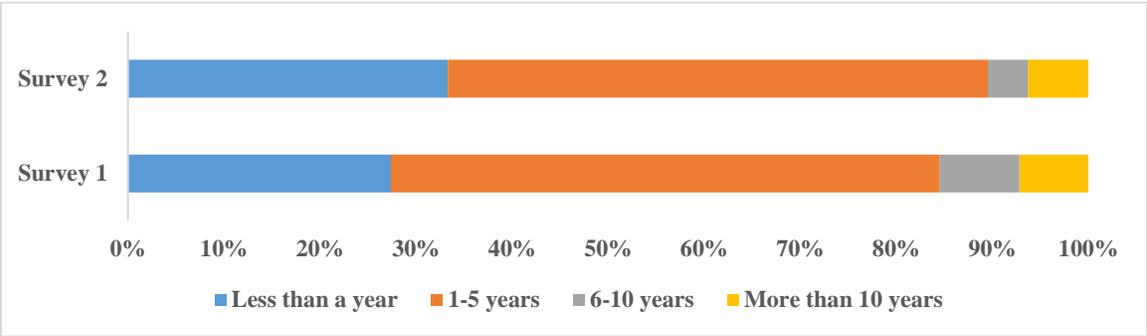

**Fig. 1.** Experience in the current company

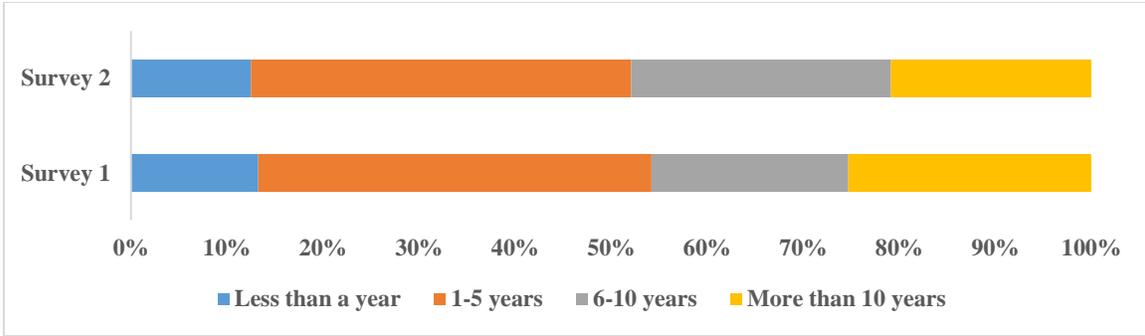

**Fig. 2.** Experience in the AEC industry

**Company Related Information**

The overall distribution of the respondents is shown in the following heat map (Fig. 3). Darker areas are places with more respondents. Among the respondents with AR/VR experience, California had the highest rate, 51%, of participation (22 out of 43 respondents with high level of AR/VR experience). After that Illinois was the second highest rate, 12% (5 out of 43 respondents with high level of AR/VR experience). The third state was New York with 9% (4 out of 43 respondents with high level of AR/VR experience).

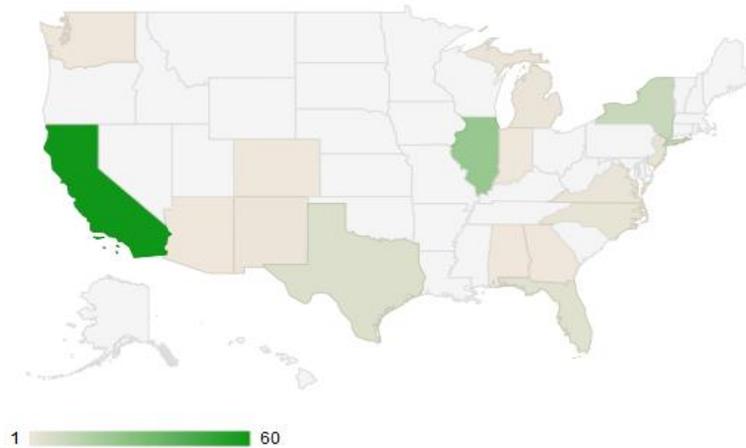

**Fig. 3.** Companies' locations

The numbers of employees and project values can be used to infer the size of a company, which can help determine how companies with different sizes envision the future of the AR/VR technologies. As Table 33 demonstrates, 17% of overall respondents were currently working at companies with more than 5000 employees (14% in survey 1 and 21% in survey 2), 26% were in 1000-5000 employees company (23% survey 1 and 32% survey 2), 21% were in 200-1000 employees company (23% survey 1 and 18% survey 2), and 36% were less than 200 employees company (40% survey 1 and 29% survey 2). Participants working for the AEC industry (excluding researchers) were also asked to identify what type of project(s) they were mainly involved with based on the average project cost (i.e., >$100 million, $10 - $50 million, etc.). As Table 44 shows, approximately 45% of participants were working on projects > $10 million in value and 50% on projects less than $5 million. It is important to note that participants had the option of choosing more than one answer for this question.

**Table 3.** Participants company size in terms of number of employees

| Size of Company | Survey 1 | Survey 2 | Overall |
|---|---|---|---|
| > 5000 Employees | 14% | 21% | 17% |
| 1000 – 5000 | 23% | 32% | 26% |
| 200 – 1000 | 23% | 18% | 21% |
| <200 Employees | 40% | 29% | 36% |

**Table 4.** Participants' company average project value

| Avg. project value | Survey 1 | Survey 2 | Overall |
|---|---|---|---|
| > $100 mil | 14% | 29% | 20% |
| $50 – $100 mil | 16% | 3% | 11% |
| $10 – $50 mil | 14% | 13% | 13% |
| $5 – $10 mil | 4% | 6% | 5% |
| $1 – $5 mil | 14% | 13% | 13% |
| $0.1 - $1 mil | 14% | 16% | 15% |
| < $0.1 mil | 25% | 19% | 23% |

The respondents had a wide variety of project types. The project types are divided into five different sectors, such as residential, commercial, institutional, transportation, and industrial (Fig. 4). Approximately 60% of the participants indicated they are involved with vertical projects and 15% working on horizontal projects. Combining the result from this question and other questions (i.e., the number of VR experts), can demonstrate the growth and adoption of AR/VR technologies in these sectors.

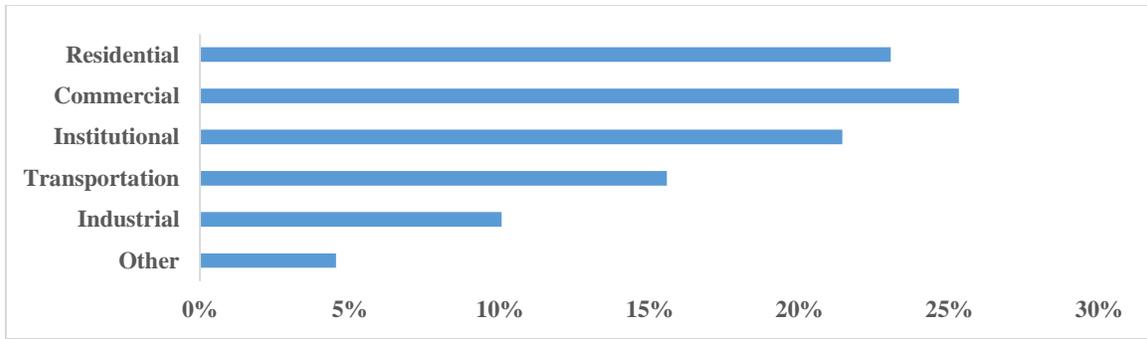
**Fig. 4.** Companies project types

## BIM Knowledge and Experience

To assess BIM knowledge of the respondents, several questions related to BIM were asked. The first question was about the BIM usage level. More than 75% of the respondents answered that they use BIM tools at least once a month. In addition, more than 90% of the engineers use BIM on a monthly basis. High usage of BIM among engineers demonstrates the importance of this technology for the industry. Fig. 5 shows the BIM usage rate for the respondents.

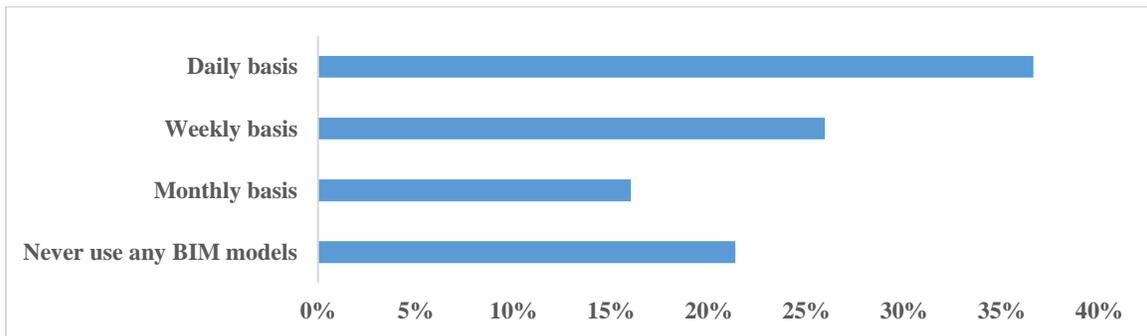
**Fig. 5.** Frequency of BIM tool usage

The second question in this section was about the experience of the respondents with BIM tools. 86% of the respondents expressed that they have had some experience with BIM tools and only 14% of the respondents have never used any BIM tool at all. Among the respondents with no BIM experience, 63% were engineers, 16% were managers, and 22% were researchers. The results show, although AEC research strongly recommends BIM, still many engineers have not used and were never trained to use any BIM tools. Fig. 6 presents respondents' experience with BIM tools.

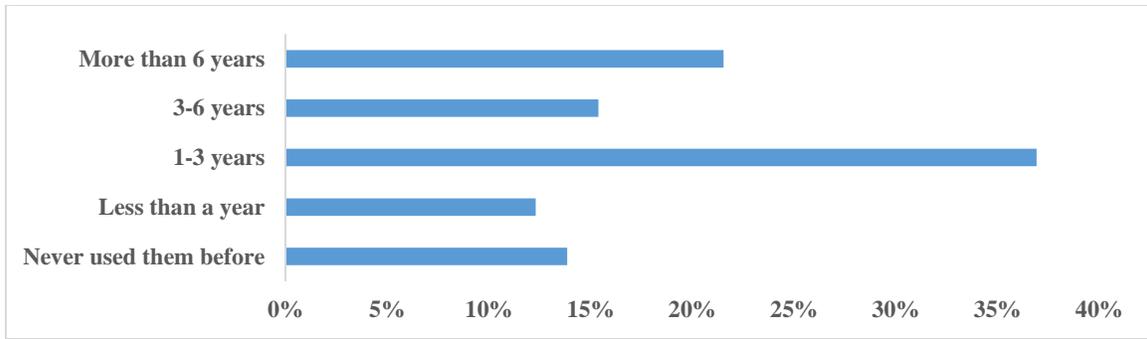

**Fig. 6.** Familiarity with BIM tools based on years of experience

The last question in this section was about applications of BIM used by the respondents. The top three applications of BIM were clash detection, model validation, and visualization and trade coordination. Using BIM for facility management purposes, energy and light simulations, transportation, and cost estimation were the least options that were chosen by the respondents. Although there were several BIM tools available in the aforementioned areas, the adoptions of BIM tools in these areas were significantly lower as shown in Fig. 7. However, the deficiency of BIM in these areas means more room for potential applications of AR/VR technologies.

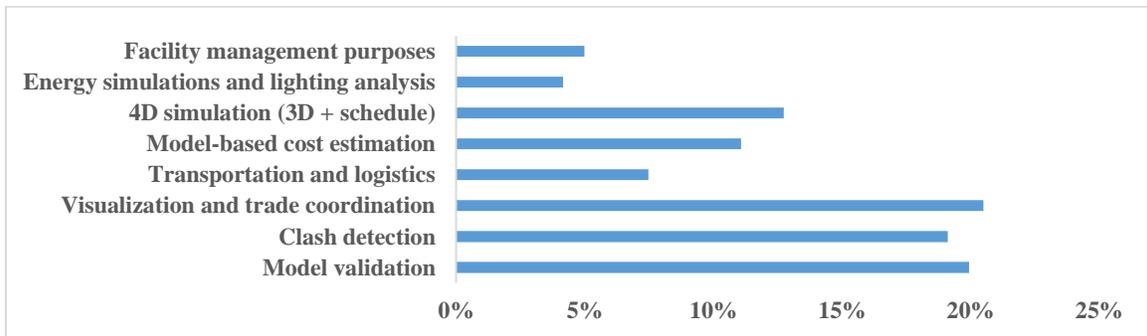

**Fig. 7.** Main sections that the respondents used BIM

## AR/VR Knowledge and Experience

This section evaluates the adoption of AR/VR technologies in the AEC industry over the past year by comparing the result of the first round of the survey with the result of the second round. In each survey, the respondents were asked about their familiarity with AR/VR equipment and whether they have used any related tools. As shown in Fig. 8, there has been a significant increase in respondents' familiarity and use of AR/VR tools from the first survey to the second survey. This growth indicates that companies and AEC professionals are becoming more familiar and interested in adopting AR/VR tools.

Fig. 8 shows how many of the respondents used AR/VR devices and applications in general. The survey made it clear that in this question, AR/VR is not only for the AEC industry, however, the survey made it clear that the "use" is only for the AEC industry in Figures 13, 14 and 15 further represents how these tools are being or will be used with the AEC industry

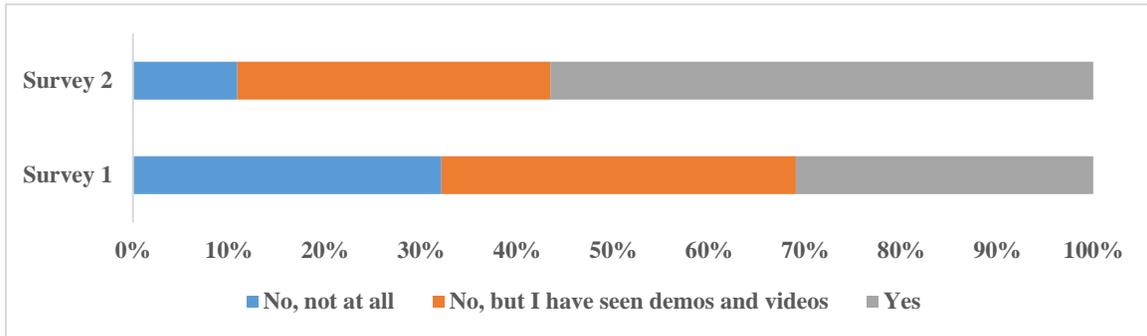

**Fig. 8.** Usage of AR/VR

Fig. 9 indicates respondents' self-reported expertise and level of understanding of AR/VR technologies. The collected data shows 5% and 13% increase in the "extremely well" and "very well" expertise and understanding categories, respectively, between the two surveys. This growth indicates there has been a significant increase in integration of AR/VR tools within AEC projects, where industry professionals are more exposed to these tools and have a better understanding of their capabilities.

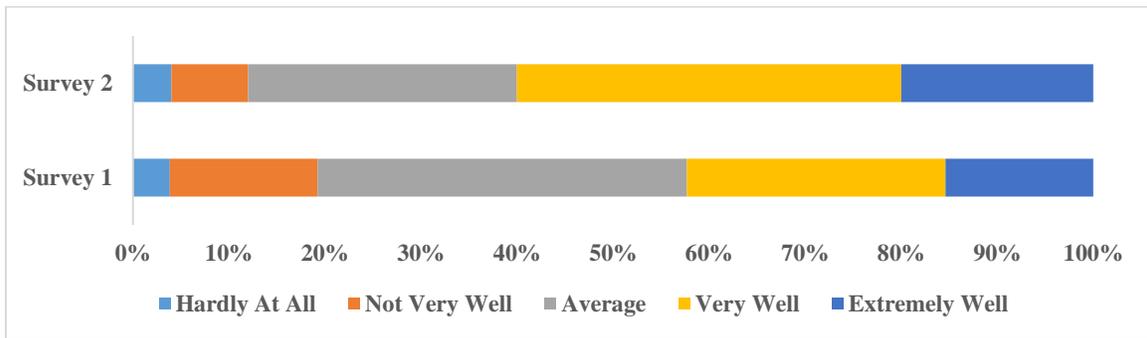

**Fig. 9.** Understanding and expertise in AR/VR tools

Respondents were also asked about which VR devices they are more familiar with and recommend to be used. The results of both surveys indicate that respondents are most familiar with and recommend Oculus Rift (approximately 45%), followed by HTC Vive, Samsung Gear, and Microsoft HoloLens. Comparing the results of first and second surveys, respondents' significantly increased recommending the use of HTC Vive as well as slight increase in Microsoft HoloLens. Consequently, recommendations for Oculus Rift and Samsung Gear marginally decreased. Fig. 10 presents the respondents' recommendation for AR/VR devices.

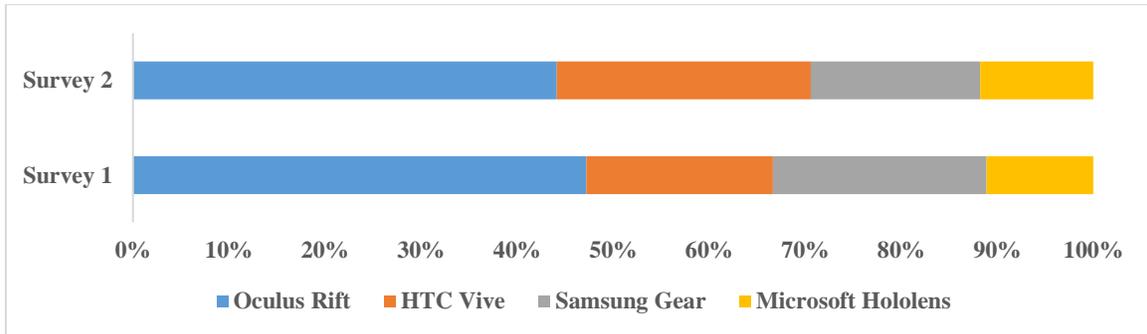

**Fig. 10.** Recommendation for VR device

The last question in this section is about the number of the AR/VR experts in the respondents' companies. As it is shown in Fig. 11, more employees are becoming familiar with AR/VR tools among the respondents' companies. This result may also indicate that the industry is adopting AR/VR technologies at a fast pace.

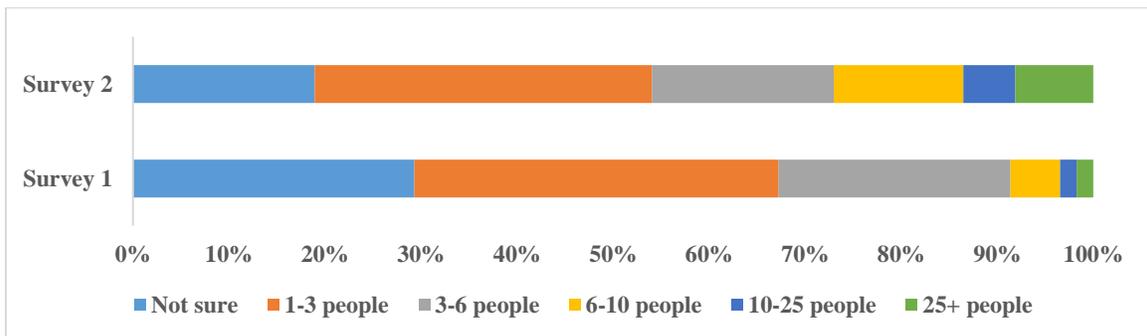

**Fig. 11.** Number of AR/VR experts in the company

## Visions of the Future AR/VR

This section was designed to determine the opportunities of AR/VR in the AEC industry. Respondents were asked to predict whether AR/VR will be used on all or majority of the projects within the next 5 to 10 years. More than 70% of all respondents chose "probably yes" or "definitely yes," indicating a significant increase in the adoption of AR/VR technologies. In addition, over the past year, the percentage of "definitely yes" and "probably yes" increased by 14%, indicating a rapid and positive change in the industry trend. Fig. 12 presents respondents' predictions on the AR/VR usage in the AEC industry for the next 5 to 10 years.

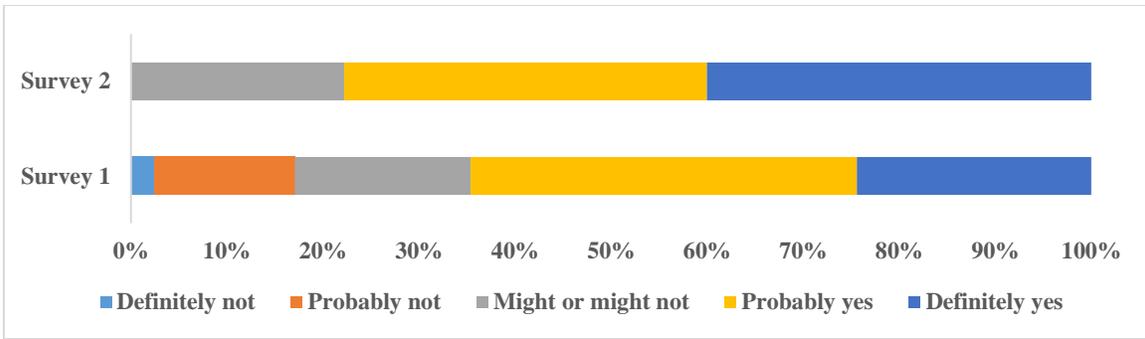

**Fig. 12.** Beliefs about AR/VR usage on all/majority of the projects within 5 to 10 years

The respondents were also asked to identify the sector that has the highest potential for the growth in VR utilization. Most of the sectors had the same rate, but the result shows that the healthcare facilities and commercial buildings are more promising. Fig. 13 presents this result.

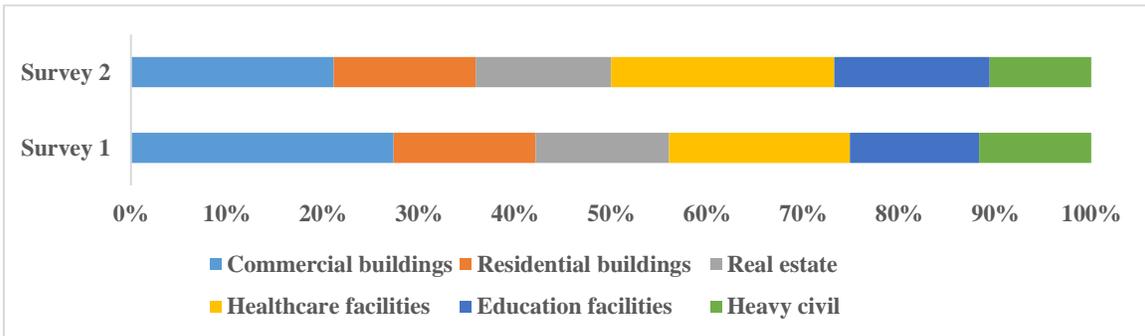

**Fig. 13.** The sectors corresponding to the most growth in AR/VR

The last question of this section asked for an optimal project size to which AR/VR can be most beneficial. Large projects had the highest response, showing that large and mega projects can make the most out of AR/VR technologies compared to small and medium projects. Fig. 14 presents this result.

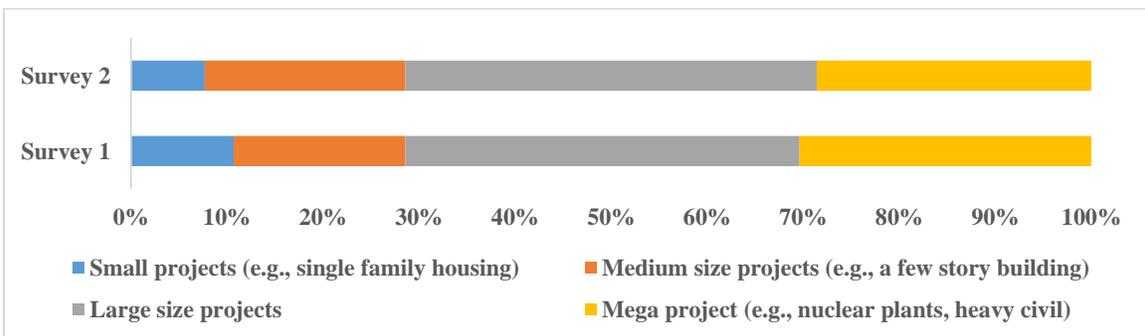

**Fig. 14.** Project sizes that achieve the highest benefits from AR/VR

In the last section of the survey, the main opportunities and limitations of AR/VR were questioned. Respondents were asked to estimate the increase in end-user (i.e., owners, contractors, and occupants) satisfaction. Approximately 90% of agreed that AR/VR can either "significantly" or "somewhat" improve the customer satisfaction rate. Furthermore, there was a growth in positive answers, from the first round of the survey to the second, as shown in Fig. 15.

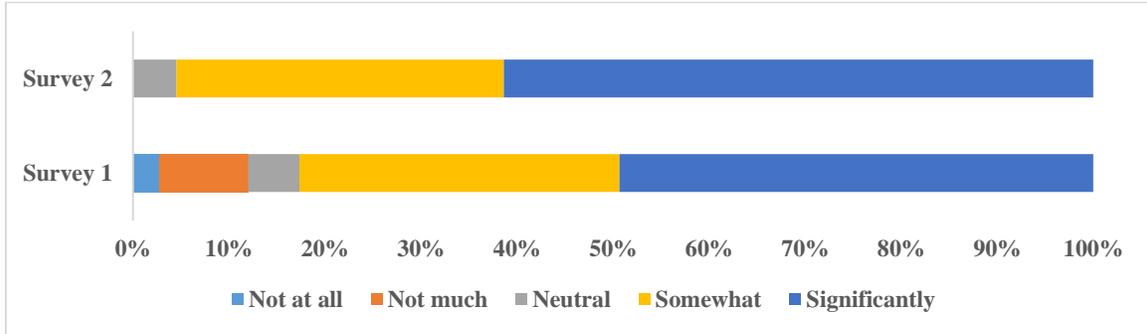

**Fig. 15.** Increase in end-users satisfaction rate by integrating AR/VR

The respondents were also asked to identify the limitations of AR/VR technologies; 21% of the respondents mentioned that lack of budget, 17% indicated upper management's lack of understanding of these technologies, and 17% of the respondents mentioned design teams' lack of knowledge as the main limitation for AR/VR utilization. Fig. 16 presents this result. Addressing these limitations can further increase the adoption of AR/VR technologies in the AEC industry.

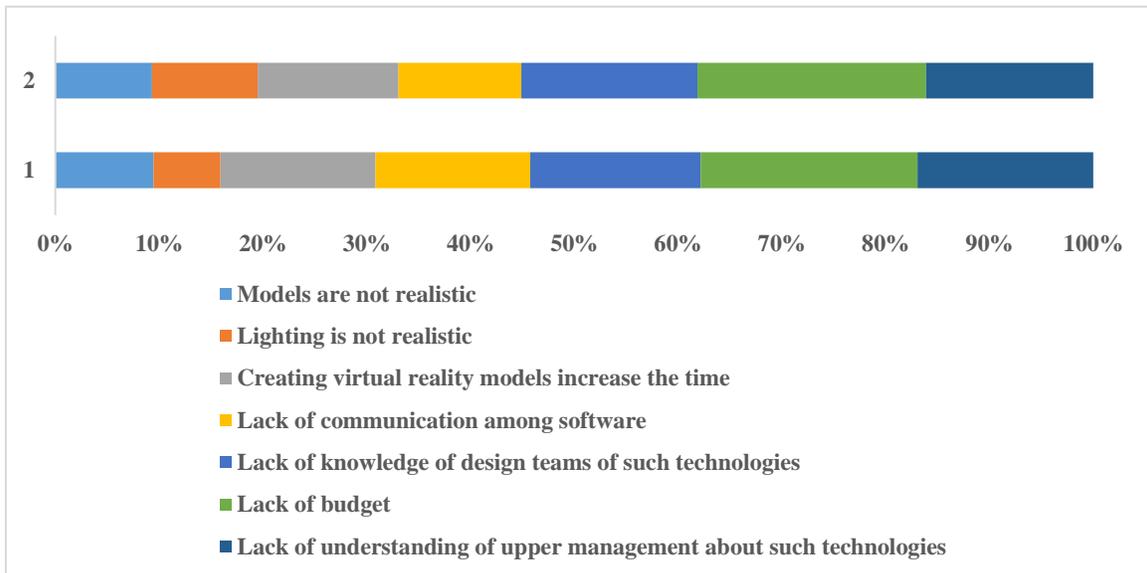

**Fig. 16.** Main limitations for AR/VR in the AEC industry

The last question of this section asked respondents for their estimate of time and cost savings (if any) in different phases of a project by adopting AR/VR technologies. The respondents' options for this question were based on the savings in terms of the project cost percentage. Approximately 55% of the respondents predicted more than 1% savings can be achieved by integrating VR/AR tools during the design and construction phases. Over 60% predicted savings of 1% during the operation phase. Fig. 17 shows the result in the design and construction phases. Fig. 18 shows the operation phase.

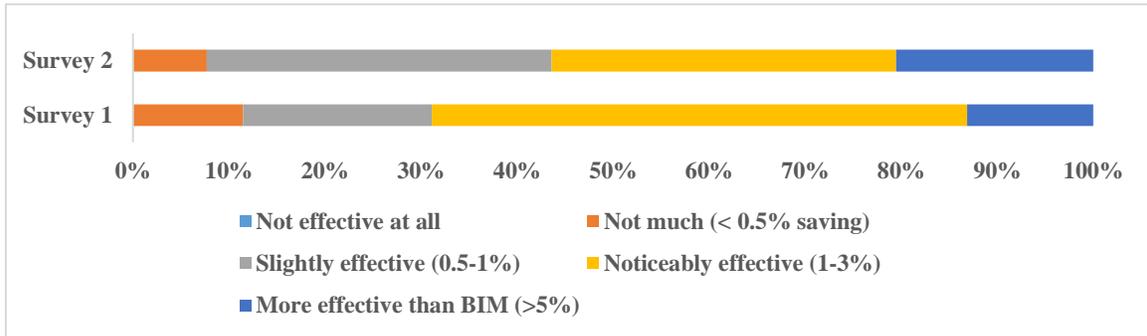

**Fig. 17.** Cost and time savings by utilizing AR/VR during the design and construction phase

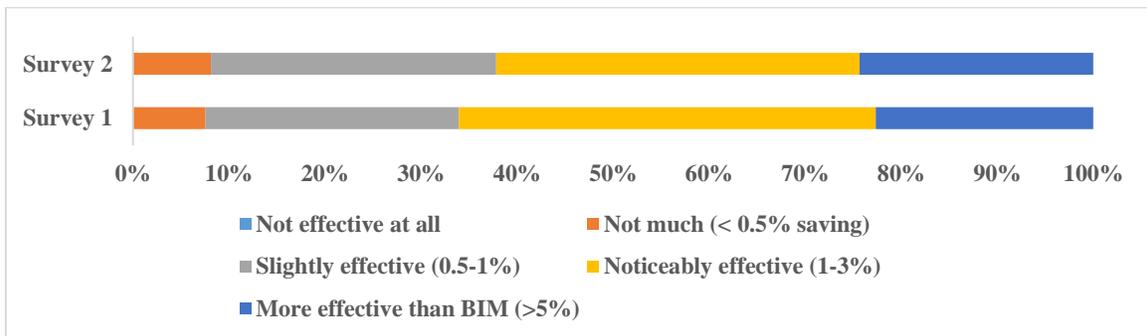

**Fig. 18.** Costs and time savings by utilizing AR/VR during the operation phase

## Discussion and Analysis

This section further discusses the survey results and how the results were analyzed. The main softwares that the author used were IBM SPSS and Excel. Furthermore, the authors mainly used unpaired t-test hypothesis testing in this section.

To measure the growth of confidence level of the respondents, the respondents' prediction on whether or not AR/VR technologies will be used on the majority of the projects within the next 5 to 10 years was analyzed. The result from unpaired t-test indicates that there was a significant difference in the scores (definitely not=0, probably not=1, might or might not=2, probably yes=3, definitely yes=4) of this question for first survey (M=2.63, SD=1.13) and second survey (M=3.20, SD=0.76); p = 0.001. These results suggest that the confidence level of respondents about the future of AR/VR technologies in the second survey are significantly higher than respondents in the first survey. This means that the AEC experts are paying more attention toward the AR/VR technologies. The increase in the number of employees with some level of expertise in AR/VR technologies between the two surveys also supports this finding.

In addition, although it seems that respondents who are relatively younger (i.e., less than 35 years old) believe that AR/VR technologies will be used on the majority of the projects within the next 5 to 10 years, the survey results indicate that older generations are more confident about the future of these technologies. The number of participants who are relatively young (less than 35 years old) is 78 and the participants from the older generations had a total number of 36. The rest of the participants were not interested in sharing their age so the authors ignored their answers for this analysis. These numbers shows that the authors had large enough samples to assume that the distribution of the data was normal. An unpaired t-test was conducted to compare younger (younger than 35 years old) and older (older than 35 years old) generations' idea about the future of AR/VR. There was a significant difference in the scores for younger generations ($M=2.86$, $SD=1.01$) and older generations ($M=3.29$, $SD=0.77$); $p = 0.025$. These results suggest that older generations' positive beliefs about the future of AR/VR technologies are significantly higher than younger generations.

Moreover, the increase in the number of employees with some level of AR/VR expertise indicates the growth in utilization of such technologies. Performing unpaired t-test on survey data shows that there was a significant difference in the number of employees with some levels of AR/VR expertise between the first survey (*M=1.24, SD=2.99*) and the second survey (*M=3.55, SD=0.65*); *p = 0.015*. These results suggest that there was a significant increase in employees becoming familiar with these technologies over the past year. Moreover, the authors used unpaired t-test and investigated the growth in employees expertise and there was a significant difference in the scores (hardly at all=0, not very well=1, average=2, very well=3, extremely well=4) of AR/VR for the first survey (*M=0.64, SD=1.18*) and the second survey (*M=1.27, SD=1.51*); *p = 0.009*. These results demonstrate that AR/VR related expertise of the respondents in the second survey was significantly improved compared to the first survey. Using unpaired t-test and dividing the companies into two groups of small companies (less than 1000 employees) and big companies (more than 1000 employees) shows that there was a significant difference between the number of AR/VR experts for the small companies (*M=1.44*, *SD=3.69*) and the big companies (*M=2.81, SD=5.56*); *p = 0.07*. This suggests that bigger companies are using these technologies more than small companies.

To show the prediction of AEC experts about the potential savings of AR/VR, the authors performed an unpaired t-test on the results from the last two question of the survey. Unpaired t-test demonstrates that there was a significant difference in the potential cost and time savings score (percentage of entire project value) in design, construction, and operation by utilizing AR/VR. The results for the first survey (M=3.21, SD=6.7) and the second survey (M=4.17, SD=10.97), p = 0.049 suggest that respondents' prediction about savings through AR/VR was significantly increased in the second survey. Also, unpaired t-test shows that there was a significant difference in the potential predicted savings scores of AR/VR from respondents with no BIM experience (M=2.90, SD=2.94) and BIM experts (M=3.80, SD=2.71); p = 0.033. These results suggest that respondents with higher BIM experience predict significantly more savings through AR/VR compared to respondents with no BIM experience.

Using the survey results, lack of budget, lack of understanding of upper management about AR/VR technologies, and lack of knowledge of design teams were the top three reported limitations for utilizing AR/VR technologies. Also, an unpaired t-test on the survey results indicates that there was a significant difference in the AR/VR expertise between engineers (M=0.76, SD=1.19) and managers (M=1.34, SD=2.71); p = 0.033. These results suggest that although managers seem not experienced with AR/VR related technologies, they are significantly more experienced than the engineers. This suggests that the AEC industry needs more AR/VR experts to fill this gap and overcome the limitations of utilizing AR/VR technologies. Additionally, 15% of respondents identified the lack of communication among software platform as one of the in survey 1; however, this number was significantly reduced (by 3%) in survey 2, indicating that algorithms and plugins are being developed that have improved this limitation. Furthermore, the data does not show any more significant results by analyzing, gender, occupation, company location, and company project type.

Although AEC industry is far behind other industries such as healthcare and retail in adopting AR/VR technologies in research literature, the results of this study showed that AEC industry is changing its previous path and therefore it is utilizing these technologies more than before. In addition this study determined the main limitations for AEC industry to adopt these technologies as well as revealing the areas that show more promises for the industry. This article also presented how different parties in AEC envisions future of AR/VR technologies as well as savings that can be achieved using AR/VR technologies.

## Conclusions and Future Vision

This paper presents two rounds of a survey that were conducted at two different time periods with about a year part. The results were analyzed to assess the current state, growth, and saving opportunities for AR/VR technologies in the AEC industry. The results of the surveys show that the industry experts foresee a strong growth in the use of AR/VR technologies over the next 5 to 10 years. Furthermore, the results show a significant increase in AR/VR utilization in the AEC industry over the past year and potential opportunities.

The surveys show some inherent limitations in adopting new AR/VR technologies, such as lack of budget, upper management's lack of understanding of these technologies, design teams' lack of knowledge. Due to lower profit margins on construction projects, one major limiting factor that prevents the industry from adopting AR/VR technologies is the lack of availability of cost/benefit analysis. Owners and companies are not willing to invest their money without knowing the true costs and benefits (i.e., time and cost savings). Therefore, there is a need for empirical studies that assess the true costs of implementing these technologies and reduction in costs and time from design to operation and maintenance phases. With regards to the other two major limitations, the results show that within the one year period between the two surveys, the number of people within the respondents' companies that are familiar with AR/VR technologies has significantly increased; this may indicate that upper management and designers/engineers will become more familiar with the capabilities of these tools in the near future as these tools become more accessible to the general consumer.

Although this paper focuses on the benefits of both the AR and VR technologies, a more detailed study is required to better identify the benefits of each technology within the AEC industry. For instance, the survey results indicate that these technologies can be very effective for model visualization, validation, and clash detection, which are tasks related to pre-construction. However, with recent advancements in mobile augmented reality and machine learning, it is expected that AR head-mounted displays provide a better assistant to project teams during the construction phase (e.g., real-time safety feedback, progress monitoring) or facility managers during the operation phase (e.g., sensor data visualization, energy simulations) in comparison to VR tools.

Although respondents indicated that communication among software has improved within the past year, there still exists a number of limitations that can improve the capabilities of VR/AR technologies for AEC professionals. For instance, there is no robust approach for transferring all BIM information along with cost data into VR platform. Importing BIM models into a 3D engine is a challenge because some of the building information (i.e., material library) might be lost during the export and import process. Moreover, connecting several VR headsets to enable a group meeting in a virtual space can enhance and improve communications among stakeholders. These problems have to be solved in order to convince the AEC industry to spend more money on the development and adoption in this area.